\documentclass[a4paper,11pt]{article}
\usepackage{jinstpub} 
\usepackage{lineno}
\usepackage{xcolor}
\usepackage{graphicx}
\usepackage{amsmath}
\usepackage{siunitx}
\usepackage{subfig}
\usepackage{url}
\usepackage{hyperref}
\usepackage{cleveref}
\usepackage{xfrac}

\usepackage{lineno}

\title{
\boldmath Ultra-high precision high voltage system for PTOLEMY}
\author{ 
R.~Ammendola$^{a}$,
A.~Apponi$^{b1}$,
G.~Benato$^{c,d}$,
M.G.~Betti$^{e,f1}$,
R.~Biondi$^{c,d}$,
P.~Bos$^{g,h}$,
G.~Cavoto$^{e,f1}$,
M.~Cadeddu$^{i}$,
A.~Casale$^{k}$,
O.~Castellano$^{b1, b2}$,
E.~Celasco$^{l,m}$,
L.~Cecchini$^{e}$,
M.~Chirico$^{e,f1}$,
W.~Chung$^{n}$,
A.G.~Cocco$^{c}$,
A.P.~Colijn$^{g,h}$,
B.~Corcione$^{e,f1}$,
N.~D'Ambrosio$^{c}$,
M.~D'Incecco$^{c}$,
G.~De~Bellis$^{e,f2}$,
M.~De Deo$^{c}$,
N.~de~Groot$^{p}$,
A.~Esposito$^{e,f1}$,
M.~Farino$^{n}$,
S.~Farinon$^{m}$,
A.D.~Ferella$^{c,o}$, 
L.~Ferro$^{i,j}$, 
L.~Ficcadenti$^{e}$,
G.~Galbato~Muscio$^{e,f2}$,
S.~Gariazzo$^{q,r,a2}$,
H.~Garrone$^{q,s,t1}$,
F.~Gatti$^{l,m}$, 
G.~Korga$^{a1}$,
F.~Malnati$^{q,s,t2}$,
G.~Mangano$^{u,v}$,
L.E.~Marcucci$^{w,x}$,
C.~Mariani$^{e,f1}$,
J.~Mead$^{g,h}$,
G.~Menichetti$^{x}$,
M.~Messina$^{c}$,
E.~Monticone$^{q,s}$,
M.~Naafs$^{g}$,
V.~Narcisi$^{a, a3}$,
S.~Nagorny$^{c,d}$,
G.~Neri$^{l,m}$,
F.~Pandolfi$^{e}$,
R.~Pavarani$^{i,j}$,
C.~P\'erez~de~los~Heros$^{z}$,
O.~Pisanti$^{u,v}$,
C.~Pepe$^{s,*}$
F.M.~Pofi$^{c,d}$,
A.D.~Polosa$^{e,f1}$,
I.~Rago$^{e}$,
M.~Rajteri$^{q,s}$,
N.~Rossi$^{c}$,
S.~Ritarossi$^{b1, b2}$,
A.~Ruocco$^{b1, b2}$,
G.~Salina$^{a}$,
A.~Santucci$^{i,j,a3}$,
M.~Sestu$^{i,j}$,
A.~Tan$^{n}$,
V.~Tozzini$^{w,y}$,
C.G.~Tully$^{n}$,
I.~van~Rens$^{p}$,
F.~Virzi$^{c,o}$,
G.~Visser$^{g}$,
M.~Viviani$^{x}$
}

\affiliation{
$^{a}$INFN Sezione di Roma Tor Vergata, Roma, Italy; \\
$^{b1}$INFN Sezione di Roma Tre, Roma, Italy; \\
$^{b2}$Dipart. di Scienze, Università degli Studi di Roma Tre, Roma, Italy; \\
$^{c}$INFN Laboratori Nazionali del Gran Sasso (LNGS), L'Aquila, Italy;  \\
$^{d}$Gran Sasso Science Institute (GSSI), L'Aquila, Italy;  \\
$^{e}$INFN Sezione di Roma, Roma, Italy; \\ 
$^{f1}$Dipart. di Fisica, Sapienza Università di Roma, Roma, Italy;  \\
$^{f2}$Dipart. di Ingegneria Astronautica, Elettrica ed Energetica, Sapienza Università di Roma, Roma, Italy;  \\
$^{g}$Nationaal instituut voor subatomaire fysica (NIKHEF), Amsterdam, The Netherlands;  \\
$^{h}$Dept. of Physics, University of Amsterdam, Amsterdam, The Netherlands;  \\
$^{i}$INFN Sezione di Cagliari, Cagliari, Italy;  \\
$^{j}$Dipart. di Fisica, Università di Cagliari, Cagliari, Italy;  \\
$^{k}$Dept of Physics, Columbia University, New York, USA; \\
$^{l}$INFN Sezione di Genova, Genova, Italy;  \\
$^{m}$Dipart. di Fisica, Università di Genova, Genova, Italy; \\
$^{n}$Princeton University, Princeton NJ, USA;  \\
$^{o}$Dipart. di Fisica, Università degli Studi dell’Aquila, L’Aquila, Italy; \\
$^{p}$Dept. of Physics, Radboud University, Nijmegen, The Netherlands; \\
$^{q}$INFN Sezione di Torino, Torino, Italy;  \\
$^{r}$Dipart. di Fisica, Università di Torino, Torino, Italy;  \\
$^{s}$Istituto Nazionale di Ricerca Metrologica (INRiM), Torino, Italy;  \\
$^{t1}$Dipart. di Elettronica \& Telecomunicazioni (POLITO-ELN), Politecnico di Torino, Torino, Italy; \\
$^{t2}$Dipart. Scienza Applicata e Tecnologia, Politecnico di Torino, Torino, Italy; \\
$^{u}$INFN Sezione di Napoli, Napoli, Italy;  \\
$^{v}$Dipart. di Fisica, Università degli Studi di Napoli Federico II, Napoli, Italy; \\
\newpage
$^{w}$INFN Sezione di Pisa, Pisa, Italy; \\
$^{x}$Dipart. di Fisica, Università di Pisa, Pisa, Italy; \\
$^{y}$CNR-Instituto Nanoscienze, Pisa, Italy; \\
$^{z}$Dept. of Physics and Astronomy, Uppsala University, Uppsala, Sweden; \\
$^{a1}$Dept. of Physics,  University of Oxford, Oxford, UK; \\
$^{a2}$Instituto de Fisica Corpuscular (IFIC - CSIC/UV), Paterna (Valencia), Spain; \\
$^{a3}$Agenzia nazionale per le nuove tecnologie, l'energia e lo sviluppo economico sostenibile (ENEA), Frascati (Roma), Italy;\\
$^{*}$\emph{Now at} CSIC -  Institut de Microelectrònica de Barcelona (IMB-CNM); \\
}

\emailAdd{nicola.rossi@lngs.infn.it}

\abstract{The PTOLEMY project is prototyping a novel electromagnetic filter for high-precision $\beta$ spectroscopy, with the ultimate and ambitious long-term goal of detecting the cosmic neutrino background through electron capture on tritium bound to graphene. Intermediate small-scale prototypes can achieve competitive sensitivity to the effective neutrino mass, even with reduced energy resolution. To reach an energy resolution better than \SI{500}{meV} at the tritium $\beta$-spectrum endpoint of \SI{18.6}{keV}, and accounting for all uncertainties in the filtering chain, the electrode voltage must be controlled at the level of a few parts per million and monitored in real time. In this work, we present the first results obtained in this effort, using a chain of commercial ultra-high-precision voltage references, read out by precision multimeters and a \emph{field mill} device. The currently available precision on high voltage is, in the conservative case, as low as \SI{0.2}{ppm} per \SI{1}{kV} single board and $\lesssim$ \SI{50}{mV} over the \SI{10}{kV} series, presently limited by field mill read-out noise. However, assuming uncorrelated Gaussian noise extrapolation, the real precision could in principle be as low as \SI{0.05}{ppm} over \SI{20}{kV}.}

\keywords{XX}

\arxivnumber{XX.YY} 

\begin{document}
\maketitle
\flushbottom

\section*{Introduction}

The detection of the cosmic neutrino background (C$\nu$B), decoupled just one second after the Big Bang initial singularity, would provide a unique opportunity to test the standard cosmological model in the very early Universe~\cite{bib:cnb, bib:ptphys}. Although the C$\nu$B is the most abundant source of neutrinos in the Universe, its detection is extremely challenging due to the very low energy of these so-called relic neutrinos.
One of the most promising detection channels to overcome this difficulty is the decay of $\beta$-unstable elements induced by their interaction with the C$\nu$B~\cite{bib:cmm}. When a C$\nu$B neutrino interacts with a $\beta$-unstable isotope, it results in the emission of a monochromatic electron, shifted from the $\beta$-spectrum end-point by an energy interval equal to twice the effective neutrino mass. Consequently, any attempt to detect C$\nu$B features inherently requires extremely high sensitivity to the effective neutrino mass, a key missing parameter of the Standard Model that has yet to be experimentally determined. The current best limit in this direction has been set by the KATRIN Collaboration, which reports $m_{\nu_e}^{\rm eff} <$\SI{450}{meV} (90\% CL) with an energy resolution of approximately \SI{1}{eV} on the $\beta$-spectrum~\cite{bib:katrin}.

The PTOLEMY project is exploring the use of tritium atoms ($^3$H) bound  to graphene layers (\emph{tritiated graphene}) to enhance both the stability and efficiency of the target~\cite{,bib:concept, bib:proposal}. Based on indirect constraints~\cite{bib:desi}, the expected effective neutrino mass is likely below \SI{200}{meV}, requiring extremely high energy resolution for C$\nu$B detection. However, for effective neutrino mass measurement, a comparatively lower resolution of $\lesssim$\SI{500}{meV} would still be competitive with the current generation experiments~\cite{bib:ptphys}. To achieve this goal, the PTOLEMY Collaboration is proposing a novel electromagnetic filter composed of multiple modules, each of which must contribute minimally to the overall energy resolution. In particular, the high-voltage (HV) electrodes must maintain accuracy within a few parts per million (ppm) over about \SI{20}{kV}.

The aim of this work is to describe the HV control and stabilization approach investigated by the Collaboration and to present the first direct measurements. In Sec.~\ref{sec:demo}, a detailed description of the PTOLEMY demonstrator and its modules is provided. In Sec.~\ref{sec:hv}, the HV stability system adopted by the Collaboration is thoroughly discussed. In Sec.~\ref{sec:res}, the first results on HV stability are presented and analyzed. Finally, in Sec.~\ref{sec:fast}, the concept of a fast electron switch, required for the current baseline PTOLEMY filter, is proposed.

\section{The PTOLEMY Demonstrator}
\label{sec:demo}

\begin{figure}
    \centering
    \includegraphics[width=0.95\linewidth]{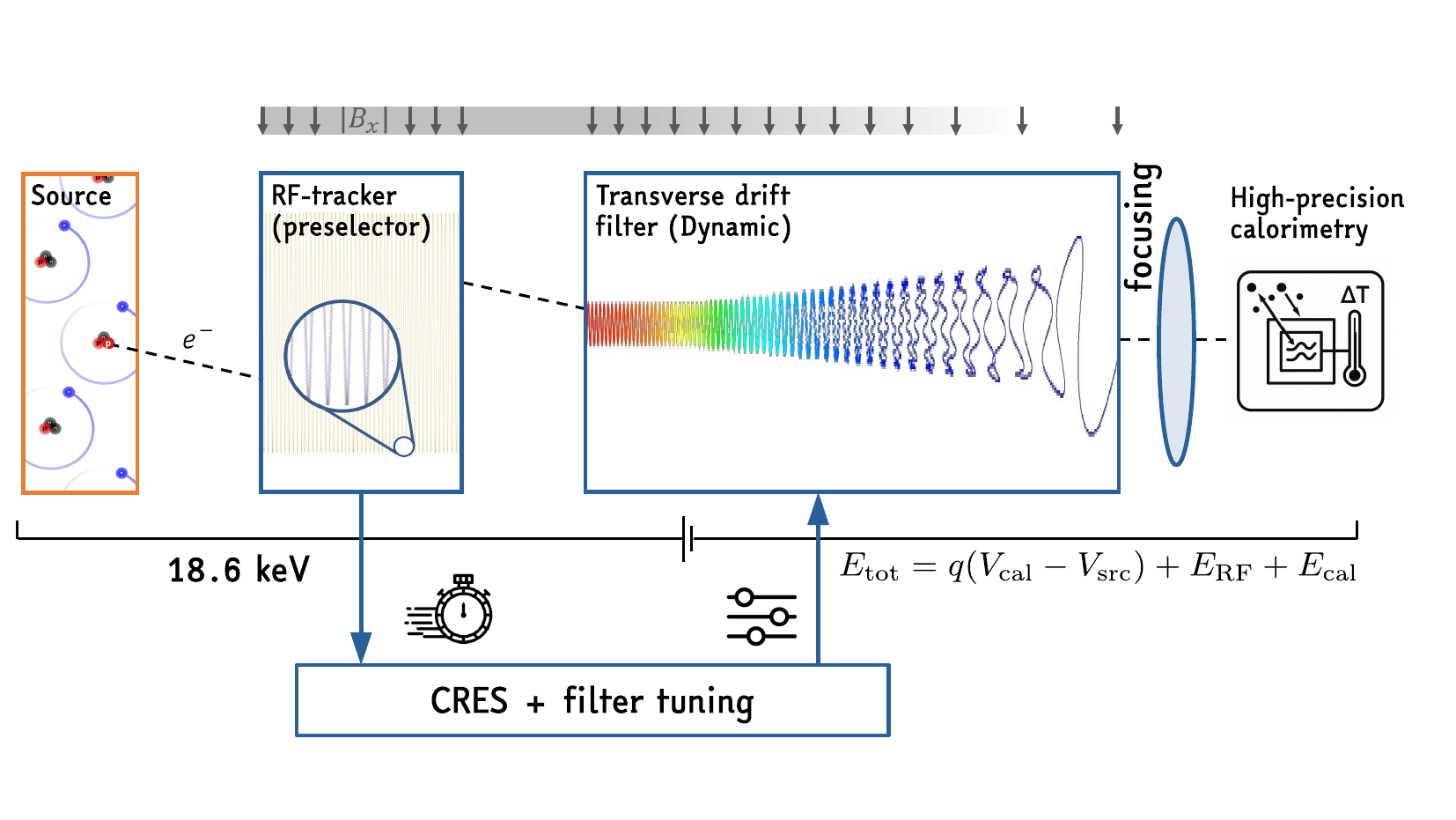}
    \caption{Schematic of the PTOLEMY concept. From left to right: tritiated graphene target, RF detection antenna (CRES) and fast trigger, dynamic electromagnetic filter and high-precision calorimeter.}
    \label{fig:demo}
\end{figure}

The goal of the PTOLEMY Demonstrator, which is being built at Laboratori Nazionali del Gran Sasso (INFN)~\cite{bib:ptweb}, is to prove the viability of the PTOLEMY detection concept~\cite{bib:concept, bib:proposal} on a reduced scale, exploring the criticalities of its individual modules both separately and in combination. Once all technical aspects are assessed, the Demonstrator can be rearranged for a real experiment to explore the neutrino mass with a small target of $1\div100$ \SI{}{\micro g} of tritium, preliminarily addressed in ~\cite{bib:ptphys}. As schematically depicted in the flow chart reported in Fig.~\ref{fig:demo}, the PTOLEMY concept consists of the following four modules.

(i) \emph{Tritiated graphene target:} tritium, featuring a low Q-value (\SI{18.6}{eV}), useful half-life (\SI{12.3}{y}), simple nuclear structure, and a relatively high cross section for neutrino capture ($3.8\times 10^{-45}$~\unit{cm^2}), is one of the best candidates for the physics investigated by PTOLEMY. It is possible to load it onto a graphene layer with high efficiency and stability~\cite{bib:graphene, bib:graphstab}.

(ii) \emph{Cyclotron radiation emission spectroscopy (CRES):} electrons from the tritium $\beta$ decay are directed into a region with a high constant magnetic field (\SI{1}{T}) and a suitable electric field allowing for bouncing and $\mathbf{E} \times \mathbf{B}$ drift (\emph{RF tracker})~\cite{bib:exb}. The radio-frequency (RF) emission associated with the cyclotron motion is detected by an RF antenna, exploiting a technology similar to the one used in Project-8~\cite{bib:cres, bib:project8}. The detected RF provides information about the energy and transverse momentum of the electron.

(iii) \emph{Dynamic transverse drift electromagnetic filter:} if the RF detection indicates that the electron has an energy reasonably close to the Q-value, a \emph{fast switch} within $\simeq$\SI{1}{ms} tunes the HV in order to allow the electron to pass through the filter. Unlike KATRIN, which transforms transverse momentum into parallel momentum over a gigantic volume, the constant drift filter requires a region of the order of a meter.
This region is characterized by non uniform and mutually perpendicular magnetic and electric fields, allowing $\mathbf{E}\times \mathbf{B}$ and ${\bf B}\times \mathbf{\nabla} B$ magnetic drifts to convert kinetic energy into potential one, see details in~\cite{bib:ptfilter1,bib:ptfilter2}.

(iv) \emph{Final electron detection:} the energy of the selected electron, slowed down to a few hundreds of eV, is eventually measured in a high-precision calorimetric system. At the moment the Collaboration is considering two options: (a) microcalorimeters based on titanium-based transition-edge sensors (TES)~\cite{bib:tes} and (b) hemispherical electron energy analyzers~\cite{bib:hemis}.

As shown in Fig.~\ref{fig:demo}, the total energy of the selected electron is given by the different contributions of each module acting on the traveling particle from the source to the high-precision calorimeters. The total uncertainty, $\sigma_{\rm TOT}$, of the electron energy required for high-precision $\beta$-spectroscopy is conceptually  given by the combinations of the most important sources of uncertainties (either statistical, systematic, or related to any possible change of efficiency),

\begin{equation} \label{eq:desum}
\sigma_{\rm TOT} \approx \sigma_{\rm Source}\otimes \sigma_{\rm RF} \otimes \sigma_{\rm Filter} \otimes \sigma_{\rm Det}\, ,
\end{equation}
where $\sigma_{\rm Source}$ represents the condensed-matter model uncertainty of electrons bound on graphene~\cite{bib:heisen, bib:angelo}, $\sigma_{\rm RF}$ accounts for energy loss due to RF emission (quantifiable as $\sim$~\SI{7}{\milli eV/ \micro s}), $\sigma_{\rm Filter}$ is primarily related to the HV applied to the filter electrodes, and $\sigma_{\rm Det}$ corresponds to the energy resolution of the final high-precision calorimeter. The symbol $\otimes$ reduces to the sum in quadratures in case those quantities are uncorrelated. 

The next Section describes in detail the strategy adopted by the PTOLEMY Collaboration to minimize the uncertainty in the HV of the filter electrode, hereafter denoted by $\sigma_{\rm HV}$. It is indeed assumed that the systematic uncertainty associated with the filter is basically dominated by the uncertainty in the HV (\emph{i.e.} $\sigma_{\rm Filter} \simeq \sigma_{\rm HV}$).

\section{The precision HV system}
\label{sec:hv}

The precision of the HV, denoted as $\sigma_{\rm HV}$, is directly related to the uncertainty in the voltage, given by
\begin{equation} \label{eq:propV}
\sigma_{\rm HV} \simeq \frac{\delta V}{V} E_{\rm e},
\end{equation}
where $V$ is the total voltage applied to slow down the electron to a few hundreds of \unit{eV}, $\delta V$ is its uncertainty and $E_{\rm e}$ is the $\beta$ emission energy. Since the tritium end-point is at 18.6 keV, the total applied voltage should be on the order of 20 kV. From the simple relation in Eq.~(\ref{eq:propV}), achieving the challenging energy resolution of \SI{50}{meV} requires $\delta V/V \lesssim 2.5$ ppm.

\begin{figure}[t!]
\centering
\includegraphics[width=0.8\linewidth]{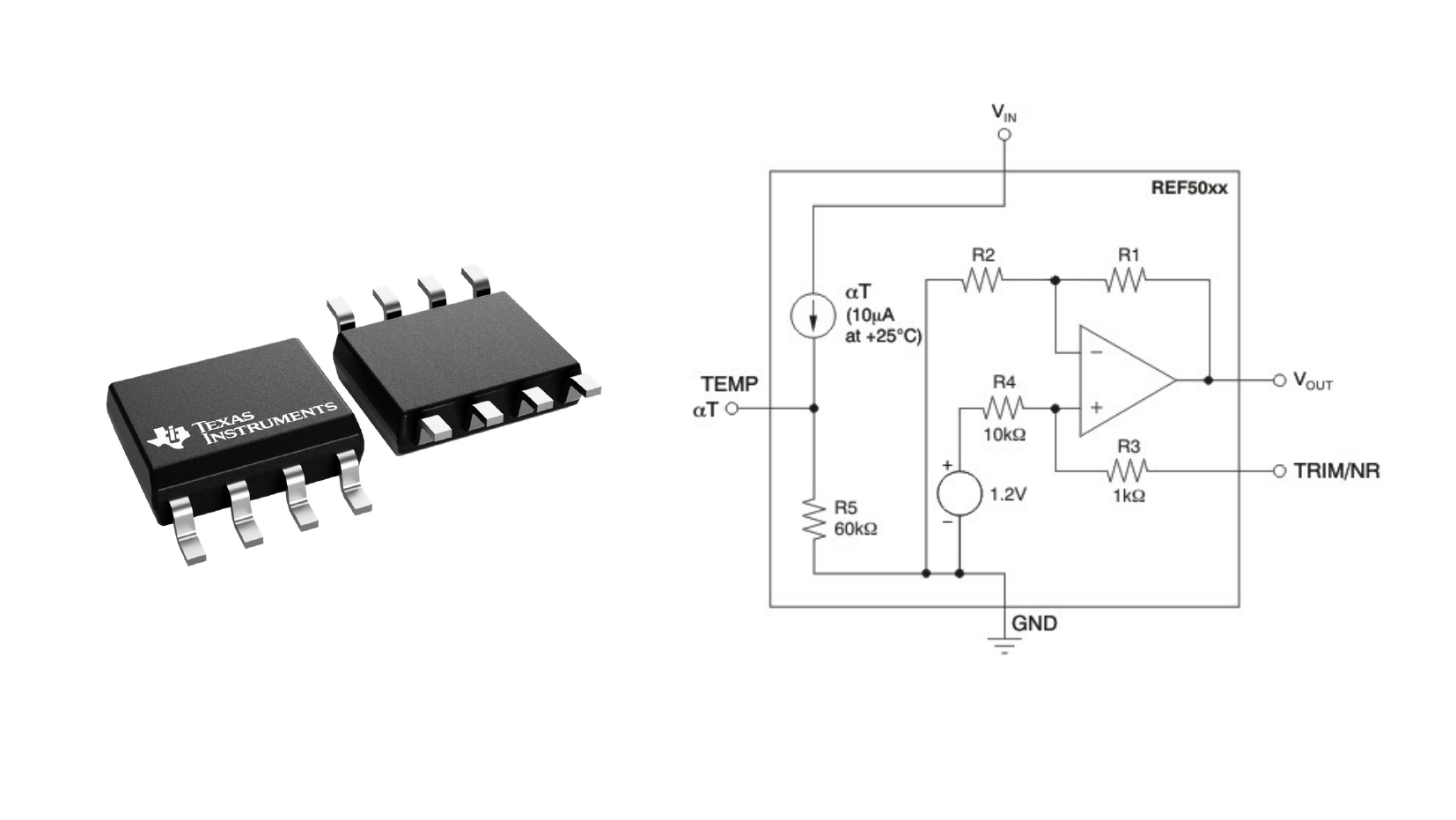}
\caption{REF5010 ultra-high precision voltage reference by \emph{Texas Instruments}. \emph{Left:} visual appearance of the electronic component. \emph{Right:} corresponding circuit diagram.}
\label{fig:ref}
\end{figure}

Before proceeding further, it is important to clarify that two aspects must be carefully controlled in the voltage reference: both \emph{precision} and \emph{accuracy}. The first refers to the consistency of measurements under unchanged conditions, while the second relates to the proximity of measurement results to the accepted value. In the following, both of these crucial aspects will be discussed. Moreover, the precision, in this sense, is the sum in quadrature of two contributions: the local fluctuations, which must be averaged, and a possible long-term component. The latter is more properly referred to as \emph{stability}. When comparing with other experiments, these two terms may have different meanings depending on the context. For example, in the KATRIN experiment, due to the integral spectrum measurement, the stability is particularly important. In contrast, in PTOLEMY, which performs a differential spectrum measurement, long-term drifts do not significantly affect the final result, as they can be accounted for in the total energy uncertainty in Eq.~(\ref{eq:desum}). However, a slow drift in the reference voltage could in principle mildly affect the filter efficiency, and this effect is under investigation by the Collaboration through a detailed filter simulation.

The strategy adopted by the Collaboration relies on an \emph{ultra-high precision} voltage reference chain (or stack)~\cite{bib:stack} and its readout, consisting of a \emph{precision multimeter} (for voltages lower than \SI{1}{\kilo V}, essential for ensuring accuracy) and a \emph{field mill} non-invasive online monitoring system (for the full \SI{20}{\kilo V} voltage, enabling final precision). 

\begin{figure}[t!]
\centering
\includegraphics[width=0.8\linewidth]{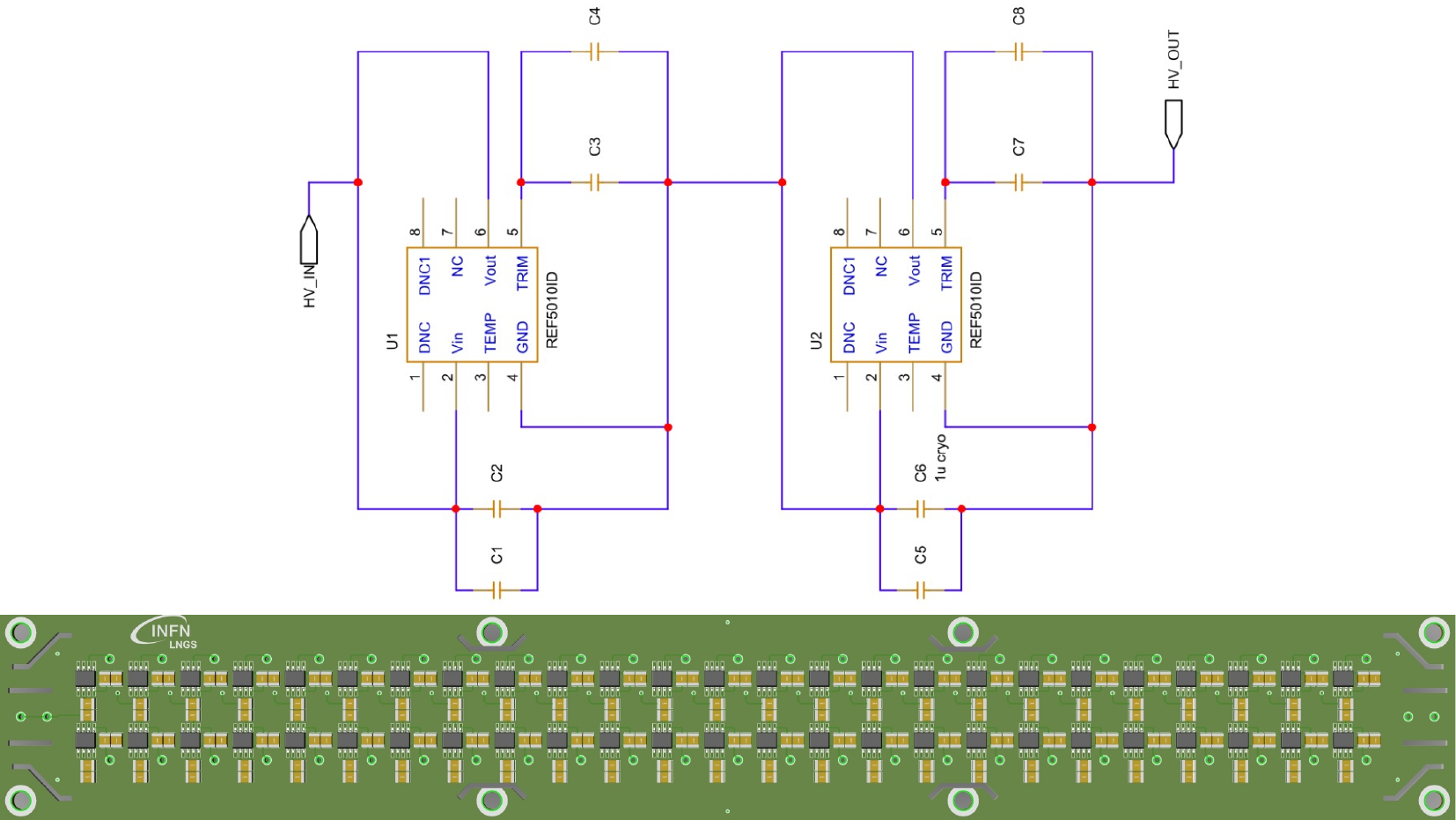}
\caption{Schematic of the repeated chain module (\emph{top}) developed at LNGS electronics laboratory. Drawing of the 1kV board (\emph{bottom}).} 
\label{fig:circuit}
\end{figure}

The basic concept behind the voltage reference chain is that, even if the single component of the voltage reference chain has a limited precision, combining a number $N$ of them under the hypothesis of Gaussian ($\sigma$) and uncorrelated noise, one can reach an ultra-high precision of the order $\sigma/\sqrt{N}$. In fact, the authors of ~\cite{bib:stack} report a precision of $5\div 7$~ppm over \SI{1}{kV} after long testings.
To a similar purpose, the KATRIN Collaboration has instead developed a high voltage divider, consisting of high precision resistors kept at constant and stable temperature and read out by a high precision multimeter~\cite{bib:weinheimer}, reaching a comparable precision. Nonetheless, after preliminary investigations, the PTOLEMY Collaboration decided to pursue the voltage reference chain testings, as they show a large profit margin, as will be discussed in this article.

The basic unit of the HV system is an ultra-high precision voltage reference, REF5010 (hereafter REF), produced by \emph{Texas Instruments}~\cite{bib:ref50xx}. Figure~\ref{fig:ref} shows the visual appearance of the electronic component (\emph{left}) and its corresponding circuit diagram (\emph{right}).
According to the data-sheet provided by the manufacturer, each REF accepts an input voltage of $V_{\rm in}=10.2\div18$ \unit{V}, returning a nominal stable output of $V_{\rm out}=10$ \unit{V} with a quiescent current of $I_{\rm q}=0.8$ \unit{\milli A}. Furthermore, the temperature drift is extremely low (2.5 ppm$/$\unit{\degreeCelsius)}, and accuracy is guaranteed at the level of 0.025\% per component. The intrinsic noise is reported as 0.5 \unit{\micro Vpp/V}. Finally, each REF features excellent long-term stability over time: 22 ppm/1000 h and operates over a wide temperature range of $-40\div125$ \unit{\degreeCelsius}.

\begin{figure}[t!]
\centering
\includegraphics[width=0.9\linewidth]{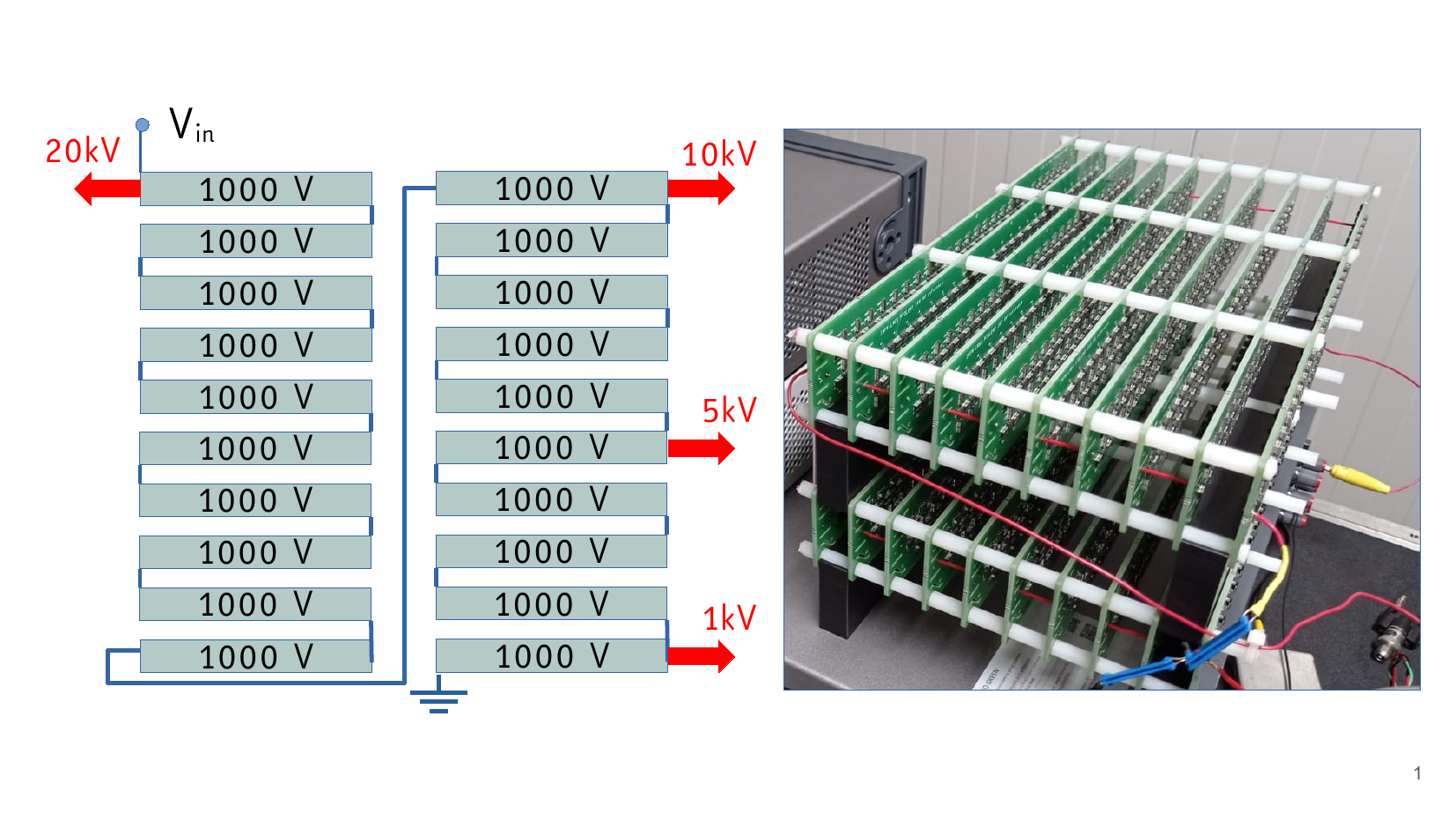}
\caption{REF Chain. \emph{Left:} schematic of the REF chain showing $V_{\rm in}$ with the four \emph{slow switch} pins. \emph{Right:} visual appearance of the twenty \SI{1}{kV} boards, arranged in two layers, as operated at LNGS during tests.} 
\label{fig:boards}
\end{figure}

As shown in Figs.~\ref{fig:circuit}~and~\ref{fig:boards}, the \SI{10}{V} REFs are arranged in series to form a long chain composed of 20 boards of \SI{1}{kV} each, summing up to \SI{20}{kV}. The boards are designed and customized in the electronics workshop of the \emph{Gran Sasso National Laboratories} (LNGS) according to the schematic depicted in Fig.~\ref{fig:circuit}. In particular, the final version, showing improved performance, was developed after extensive testing of the electronic components. The choice of specific low-noise capacitors followed a long campaign of comparative measurements. In addition, the boards were carefully designed to handle the applied high voltage and to ensure proper thermal dissipation of the components. The full chain is finally powered by a high-voltage \textsc{Bertan} 210-30R power supply. Along the chain, thanks to a \emph{slow switch} mechanical device operated by a microcontroller-driven actuator, it is possible to switch in real-time between four different voltages (namely 1, 5, 10, and 20 \unit{kV}) without physically altering the system.

\begin{figure} 
   \centering 
   \includegraphics[width=\linewidth]{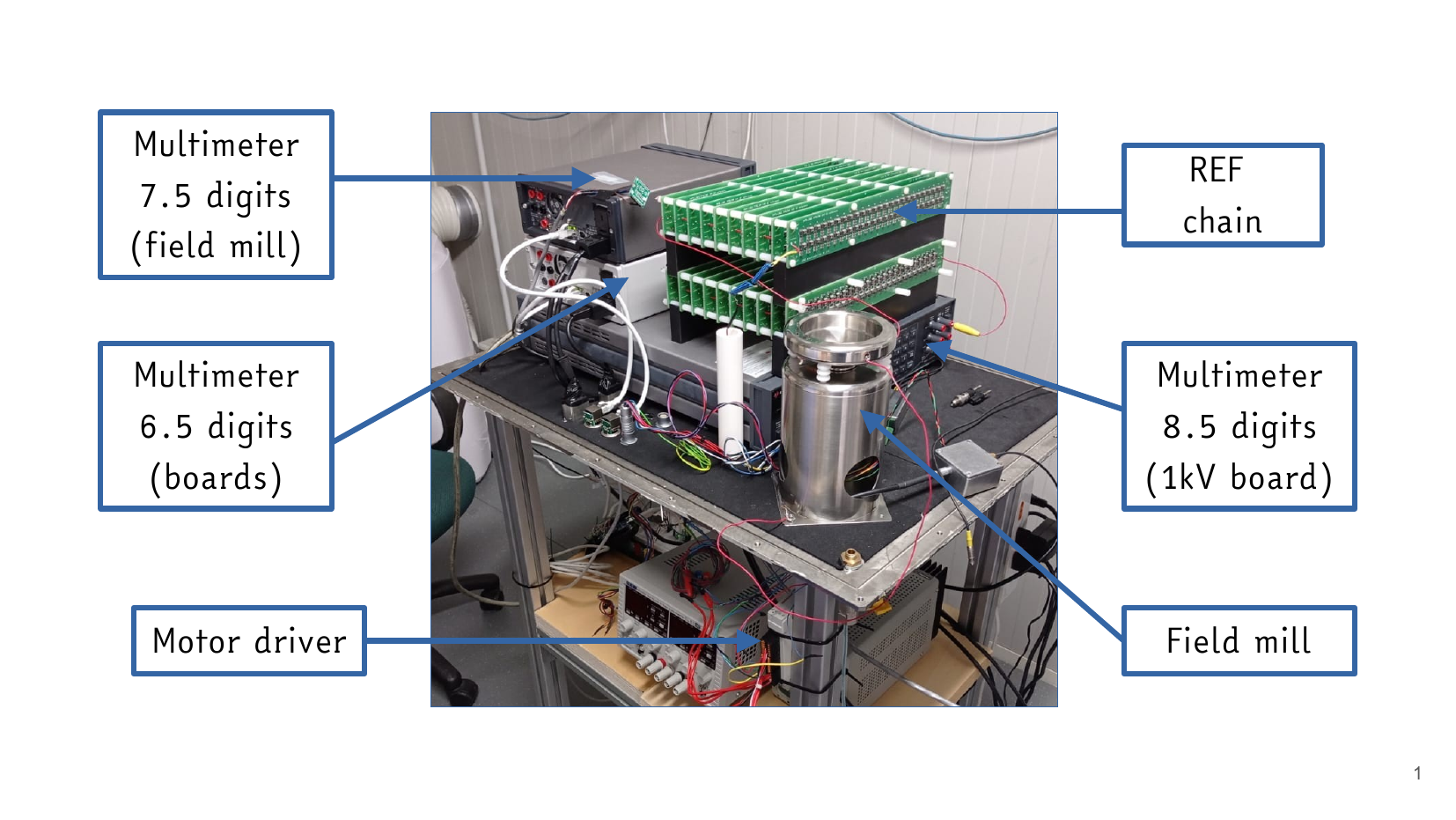} 
   \caption{Bench test of the HV system. According to the descriptions in the picture, the REF chain is monitored by an $8\sfrac12$-digit precision multimeter (for the \SI{1}{kV} board only) and by the \emph{field mill}, read out by a $7\sfrac12$-digit precision multimeter, according to the \emph{slow switch configuration}. Finally, the board temperature is read out by a 6.5-digit precision multimeter.} \label{fig:setup} 
\end{figure}

The REF board chain is installed on a test bench and connected to the readout system as illustrated in Fig.~\ref{fig:setup}. Single boards of \SI{1}{kV} REFs can be monitored by an $8\sfrac12$-digit precision multimeter (\textsc{Keysight}) 3458A for the \SI{1}{kV} corresponding to the full-scale of the device. The \emph{field mill} (see description below) accepts higher voltages up to \SI{20}{kV} according to the \emph{slow switch} configuration, and its periodic voltage signal is read out by a $7\sfrac12$-digit precision multimeter \textsc{Keysight} 34470A. Finally, a third $6\sfrac12$-digit precision multimeter \textsc{Keysight} 3446A reads out the REF temperature, produced as a calibrated low-voltage signal.

Since, as described, the operating temperature plays a crucial role in the voltage stabilization, being an important parameter in the REFs functioning, the test bench during the measurements 
is kept inside a custom made \emph{climatic chamber}. The chamber is made by a wooden box, internally coated with aluminum tape acting as a Faraday cage against environmental electromagnetic noise. The temperature is stabilized at approximately room temperature by Peltier cells, providing a total of \SI{250}{W} of cooling power, with a proportional-integral-derivative (PID) controller, optimized to compensate for the heat produced by the precision multimeters (\SI{50}{W} each) and by the REF chain (about \SI{80}{W}). The test bench inside the climatic chamber is further placed in a modular container home kept at room temperature by a \textsc{Mitsubishi} air-conditioning unit. After the PID tuning, the temperature inside the climatic chamber can be easily kept constant within a tenth of a Celsius degree, sufficient to avoid strong temperature drift related to the typical seasonal and day-night temperature variations. Finally, the chamber can be flushed with gaseous nitrogen or sulfur hexafluoride to remove humidity and minimize noise from local discharges. The gases have been flushed in many tests showing no relevant improvement on the results, for this reason most of the tests have been conducted in air.

A data acquisition (DAQ) system is capable of remotely recording all the digitized precision multimeter signals, along with different probes for monitoring environmental parameters such as temperature in various locations inside the chamber, humidity, and atmospheric pressure. All data, recorded as ASCII files, are catalogued in a database and analyzed by a custom ROOT/C++ based code.

\subsection{The \emph{field mill} read-out}

A \emph{field mill} is a device used for continuously measuring the strength of electric fields, with many applications in fields such as meteorology, space technology, and other research areas in which continuous and non-invasive electric field measurements are needed~\cite{bib:mill}. For PTOLEMY, a customized version has been designed and manufactured at the LNGS mechanical workshop.

\begin{figure}[t!] 
   \centering 
   \includegraphics[width=0.95\linewidth]{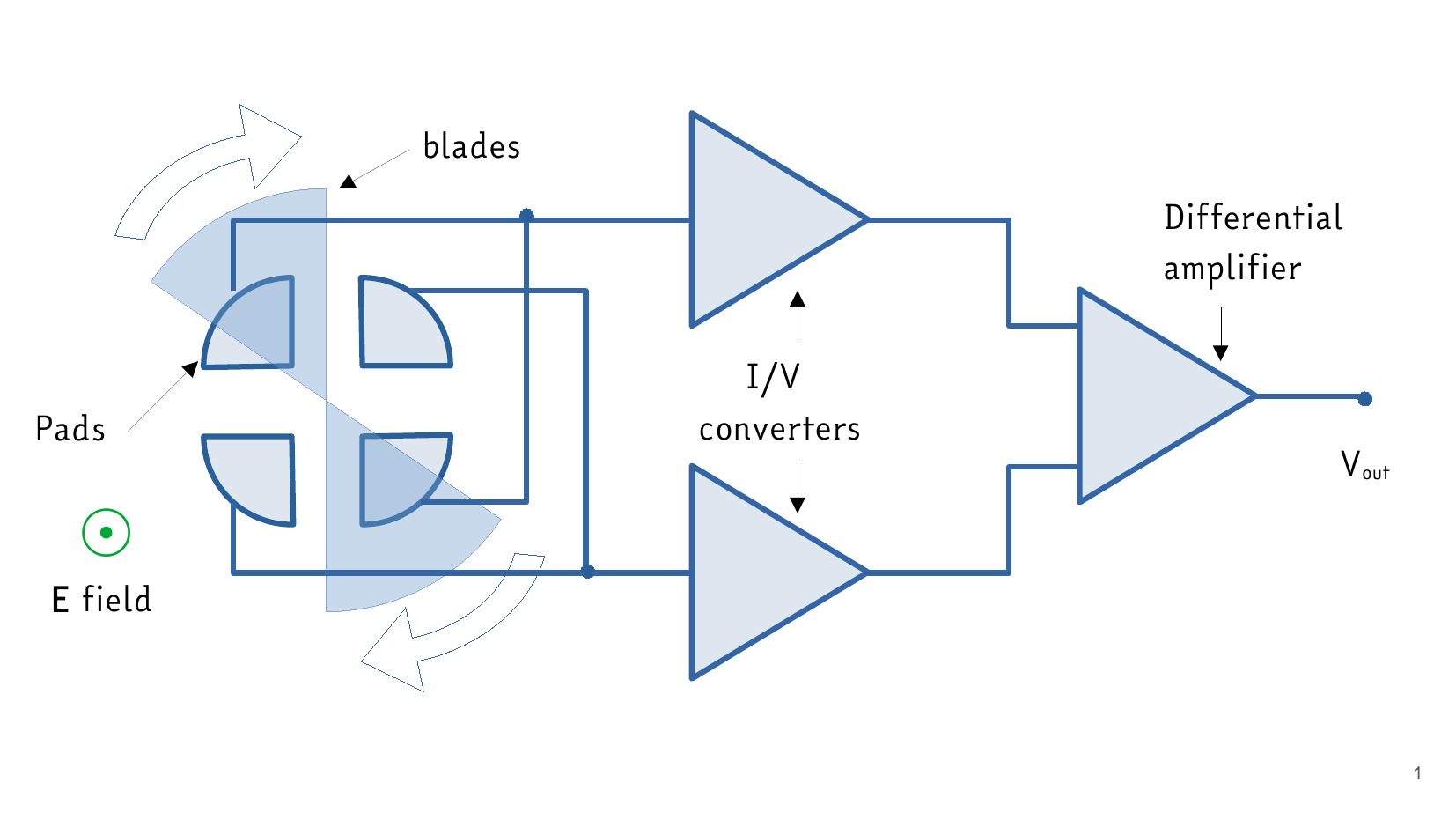} 
   \caption{\emph{Field mill} schematic.}    
   \label{fig:mill} 
\end{figure}
\begin{figure}[t!] 
   \centering 
   \includegraphics[width=0.95\linewidth]{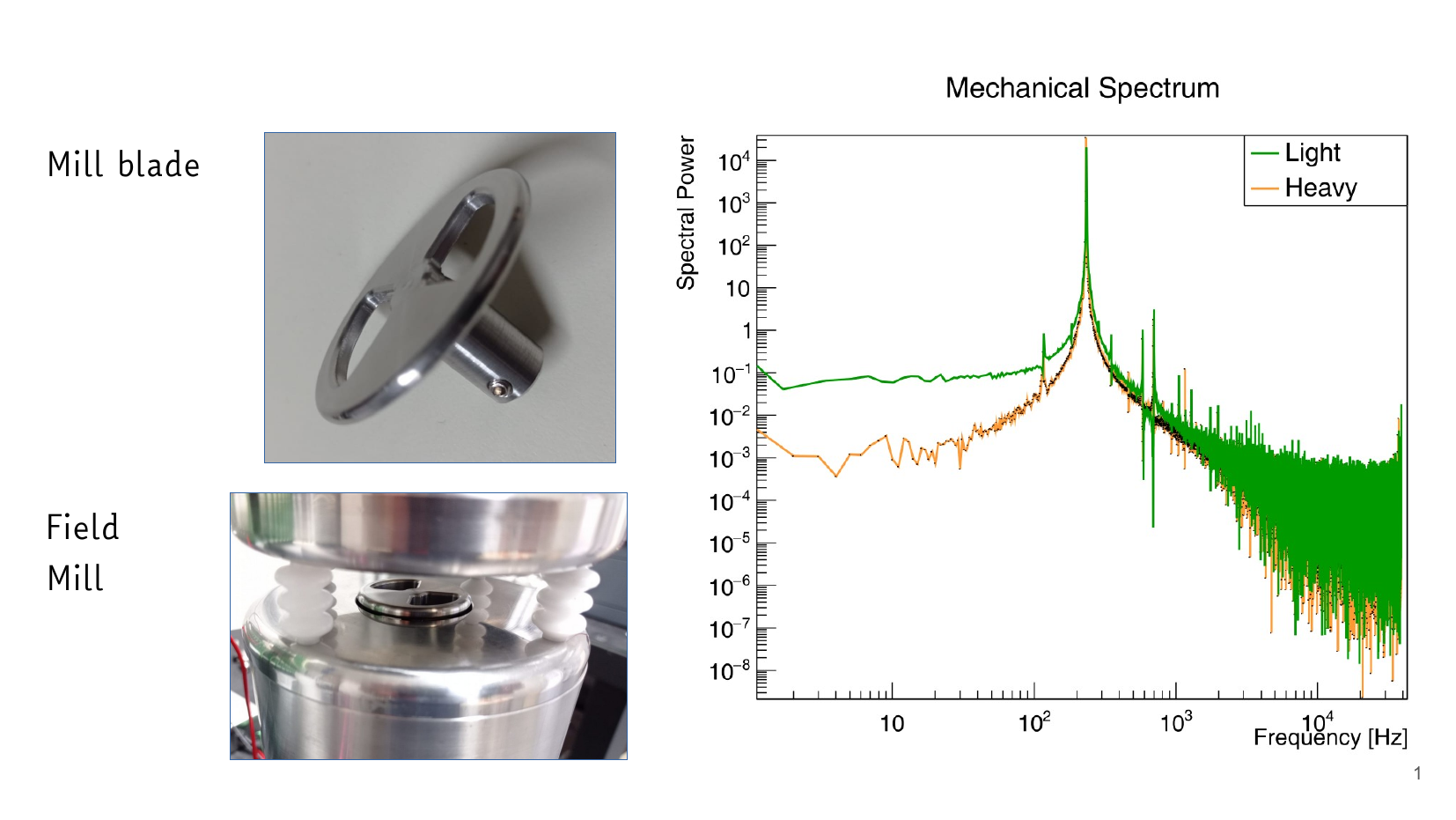} 
   \caption{\emph{Left:} picture of the mill blade (\emph{top}) and its position in between the two metal plates (\emph{bottom}). \emph{Right:} comparison between the FFT of two mill blades with different thickness. In this example the heavy version shows a better performance.}    
   \label{fig:millpics} 
\end{figure}

As shown in Fig.~\ref{fig:mill} and the detailed schematic in Fig.~\ref{fig:millpics}, the field mill consists of a top plate connected to the voltage to be measured, separated by a grounded plate through four insulating plastic spacers, shaped to minimize electrical discharges. The central part of the grounded plate, the “mill blade,” is shaped like a disk with two trapezoidal holes and rotates about the vertical axis at approximately 3000 RPM (50 Hz) by means of a brushless motor located beneath the plate.
Below the mill blade, there are four metallic pads cross-connected in pairs to a low-noise differential amplifier. The pads are electrically isolated from ground by ceramic spacers.
Due to the spinning blade, the electric field produced by the upper high-voltage plate induces an alternating charge between the two pairs of pads. The resulting amplified signal is approximately sinusoidal with a frequency of 100 Hz (the factor
two with the mill speed comes from pads connected in pair at 180°).
The signal is then readout by the high precision 8-$\sfrac12$ digit "true RMS" voltmeter. The recorded signal is proportional to the electric field produced by the voltage applied to the top plate. Therefore, once the \emph{field mill}  output is calibrated, it becomes a high-precision voltmeter.

To guarantee a high level of reliability, the blade spin must be extremely stable. For this reason, a brush-less high-precision motor activated by a driver through a PID controller has been adopted and carefully tuned. Different blade thicknesses have been tested until reaching optimal stability, quantifiable through the fast Fourier transform (FFT) of the output signal. Figure \ref{fig:millpics} \emph{right} shows an example of comparison between FTTs of the amplifier output signal for two different blade thicknesses.
Additionally, the motor speed has been optimized to maximize the signal-to-noise ratio, and low-impedance motor grounding has been carefully ensured through full-conductive metal ball bearings lubricated with conductive grease.

Except for the insulating spacers, all the parts of the \emph{field mill} are made from stainless steel. Furthermore, to minimize local electrical discharges, all sharp-edged parts have been properly rounded. In the following Section, the first results obtained with the test bench described above are reported.

\section{Results}
\label{sec:res}

\begin{figure}[h!]
    \centering
    \includegraphics[width=0.9\linewidth]{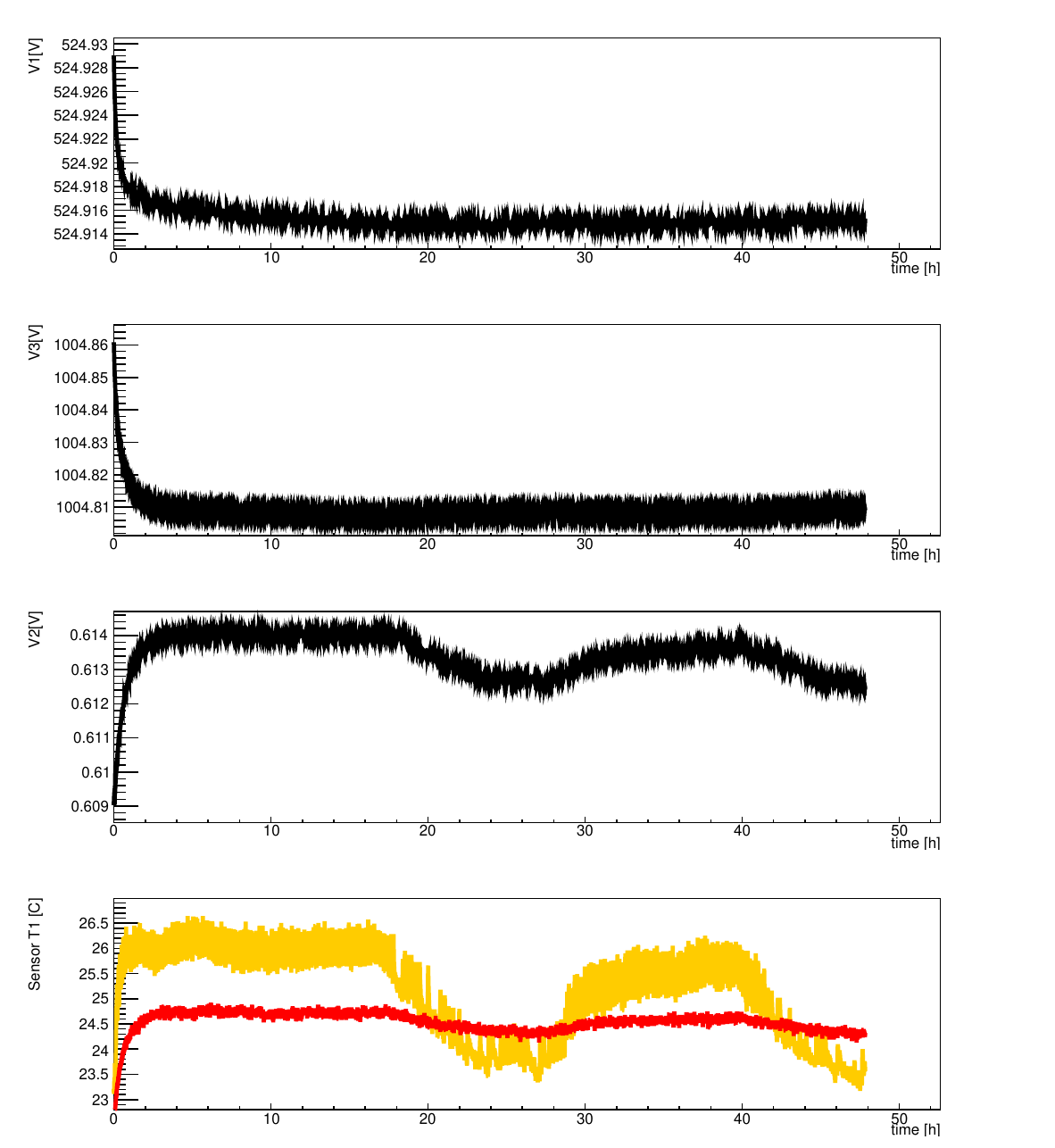}
    \caption{From the \emph{Top}: (i) voltage from 500 V pin (half-board) read out by the \textsc{Keysight}-$7\sfrac{1}{2}$, (ii) voltage from 1000 V (full-board) read out by the \textsc{Keysight}-$8\sfrac{1}{2}$, (iii) internal temperature output voltage from the first REF in the chain read out by the \textsc{Keysight}-$6\sfrac{1}{2}$ and (iv) temperature in the experimental room (\emph{yellow}) and from the climatic chamber internal sensor (\emph{red}).}
    \label{fig:trends}
\end{figure}

The signal originating from the REF boards typically consists of random Gaussian noise superimposed on long-term variations of the order of tens of minutes, if not hours. 

Figure~\ref{fig:trends} shows an example of recorded data during a 48-hour test. From the \emph{top}, the following measurements are reported:
(i) the $V_1$ voltage from the 500 V pin (half-board) read out by the \textsc{Keysight}~$7\sfrac{1}{2}$,
(ii) the $V_3$ voltage from the 1000 V pin (full-board) read out by the \textsc{Keysight}~$8\sfrac{1}{2}$,
(iii) the internal temperature output voltage $V_2$ from the first REF in the chain read out by the \textsc{Keysight}~$6\sfrac{1}{2}$, and
(iv) the temperature  ($T_1$s) in the experimental room (\emph{yellow}) and from the climatic chamber internal sensor (\emph{red}). The last two are measured by commercial sensors read out by an \textsc{Arduino} based system.
In general, the trend of the board voltages is characterized by two main time components: a fast decay lasting a couple of hours, and a long-term variation on the order of a few ppm per day, approaching an asymptotic stabilization. The temperature plot shows the accuracy of the thermal stabilization inside the climatic chamber (about \SI{0.1}{\celsius}), corresponding to stable environmental conditions in the experimental set-up. A sudden change outside the climatic chamber has a non-negligible impact on the Peltier PID control system, as visible in the second half of the time series shown in Fig.~\ref{fig:trends}. Finally, the residual temperature variations have anyway some direct effect on the board temperature, as evident from the subdominant correlation between $T_1$ and $V_2$.

These long-term variations, often correlated with slow changes in environmental conditions (primarily temperature) or caused by system relaxation, are removed through a \emph{detrending} process.
Only the short-timescale voltage fluctuations contribute to the uncertainty in the electron energy measurement.
The procedure for removing long-term variations is performed using two different methods.
The first method, used for consistency check only, involves fitting the long-term variation to a multi-component exponential curve, which is further corrected with a polynomial depending on the specific case. A second, more flexible, method, used in most of the analysis, consists of a model-independent \emph{detrend} using an optimized first-order local regression (LOESS). The sampled signal $S[n]$ is replaced at each point by the local linear regression:
\begin{equation} 
S'[n] = L_k[n, n] \, , 
\end{equation} 
where $L_k[n, n]$, for each $n$, is the center of the line $L_k[n, n+m]=\alpha_n(n+m) + \beta_n$ determined using the least squares method within the interval $ m \in [-k,+k]$, by minimizing
\begin{equation} 
    \mathbb{X}^2(\alpha_n, \beta_n) = \sum_{m=-k}^k (S[n+m]- L_k[n, n+m])^2\, . 
\end{equation} 
The parameter $k$ represents the bandwidth of the local smoothing and is determined by minimizing the mean integrated square error over the full time series:
\begin{equation} 
\min_k \left\{\sum_{n}(S'[n]-L_k[n, n])^2\right\}\, . 
\end{equation} 
This optimization ensures that the local regression follows only the long-term trend, leaving the high-frequency random component as a residual:
\begin{equation} R[n] = S[n] - S'[n].
\end{equation}

\begin{figure}
    \centering
    \includegraphics[width=1.\linewidth]{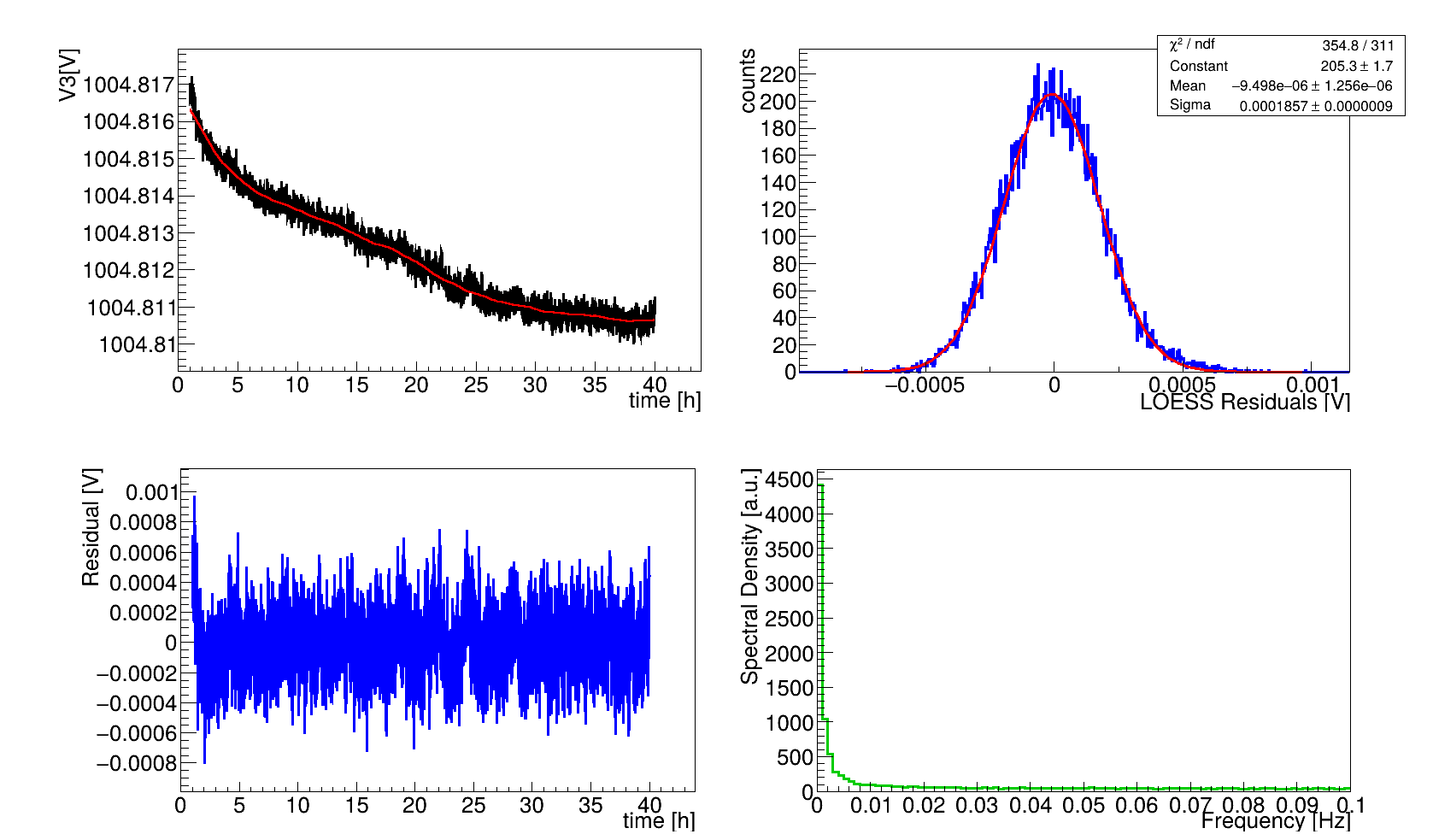}
    \caption{\emph{Top left:} REF board voltage as a function of time measured by the $8\sfrac12$-digit precision multimeter. The over-imposed \emph{red} curve represents the local regression. \emph{Top right:} Gaussian distribution of the residuals (in \emph{bottom left}) after the LOESS detrend. \emph{Bottom Right:} DFFT of the time series, showing no significant periodic components.}
    \label{fig:V1}
\end{figure}
\begin{figure} [h!]
    \centering
    \includegraphics[width=0.8\linewidth]{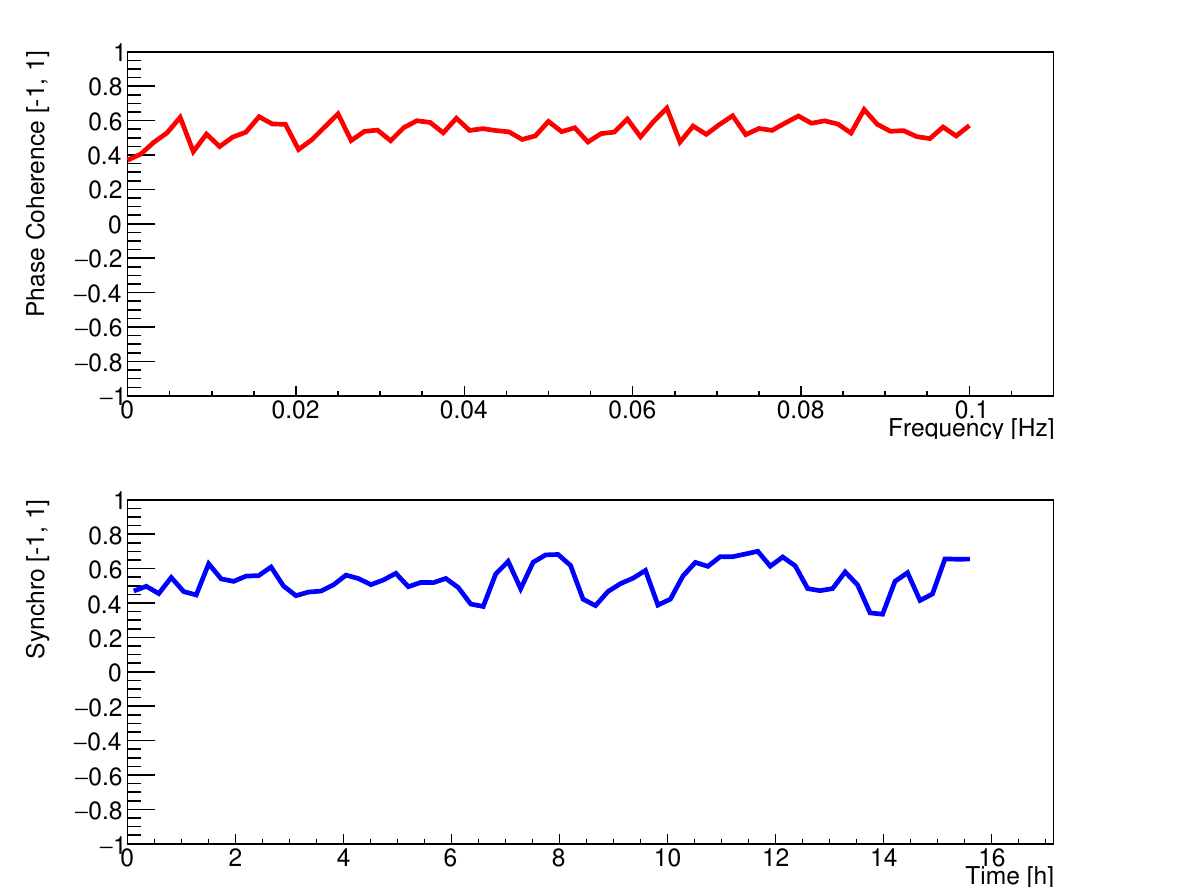}
    \caption{\emph{Top:} phase coherence in chunks of 128 sampling points between $V_{\rm 1kV}$ and $V_{\rm 0.5 kV}$. \emph{Bottom:} synchronization (covariance) between the two measurements in the same chunk. Both plots are calculated in the time interval 3-18 h with high temperature stability.}
    \label{fig:corr}
\end{figure}

Figure~\ref{fig:V1} shows an example of REF 1 kV board voltage as a function of time (about 40 h) measured by the  $8\sfrac12$-digit precision multimeter (\emph{top left}), while the over-imposed \emph{red} curve represents the corresponding LOESS detrend. As visible, the long-term variation occurs over an interval of many hours. The overall variation is about \SI{5}{\milli V}, the accuracy of the entire board, made of 100 REFs, is at the level of \SI{5}{V}, \emph{i.e.}, equal to 0.005\% and thus better than the declared technical specification of each single REF (0.025\%).   This value is compatible with the $\sqrt{100}$ reduction from the independent Gaussian uncertainty propagation, as anticipated in Sec.~\ref{sec:hv}. The precision of the single board, estimated as the RMS of the residues distribution (\emph{bottom left} and \emph{top right}), is in the worst case $\sigma\lesssim$\SI{0.2}{\milli V}, corresponding to $\lesssim 0.2$ ppm  alone (to be compared with $5\div7$ ppm reported in literature~\cite{bib:stack}). This result is far better than the minimum requirement for the final energy resolution of the PTOLEMY experiment, as described in Sec.~\ref{sec:demo}. Finally the \emph{bottom right} plot reports the discrete fast Fourier transform (DFFT) of the acquired signal, showing no significant periodic component in the time series.

The extrapolation of the precision to the full chain made of 20 boards, read out by the \emph{field mill}, can be described by the following relation

\begin{equation}\label{eq:precis} \sigma^2_{20} = \sum_{i,j}\rho_{ij}\sigma_i^2\sigma_j^2 + \sigma_{\rm mill}^2\, , 
\end{equation}
where $\rho_{ij}$ is the correlation matrix describing the extra contribution to the precision coming from possible correlated noise, and $\sigma_{\rm mill}$ is a possible intrinsic noise contribution due to the \emph{field mill} system. In the most optimistic case, if no extra noise is introduced by the readout device and no correlated noise appears in the full chain, Eq.~(\ref{eq:precis}) becomes the simple sum in quadrature of the single board noise, and the overall precision reads
\begin{equation}\label{eq:sim} 
\sigma_{20}=\sqrt{20}\,\sigma_{\rm 1 kV}  \simeq 0.9 \, {\rm mV}\, ,  
\end{equation}
resulting in the very small value of $\lesssim$ 0.05 ppm, when considered over \SI{20}{kV}.

The hypothesis of uncorrelated noise is supported by the comparison of the RMS of the trend of the full  board voltage $V_{\rm 1kV}$ and the voltage $V_{\rm 0.5kV}$ taken from the pin in the middle of the board itself at \SI{0.5}{kV}. The approximated relation $\sigma_{\rm 1 kV} \simeq \sqrt{2}\sigma_{\rm 0.5 kV}$ is basically found, as expected, from the independent uncertainty propagation. 
Another argument in favor of negligible cross correlation terms arises from the study we performed of possible \emph{phase coherence} and \emph{synchronization} between $V_{\rm 1kV}$ and $V_{\rm 0.5kV}$. 
\begin{figure} 
   \centering 
   \includegraphics[width=1.\linewidth]{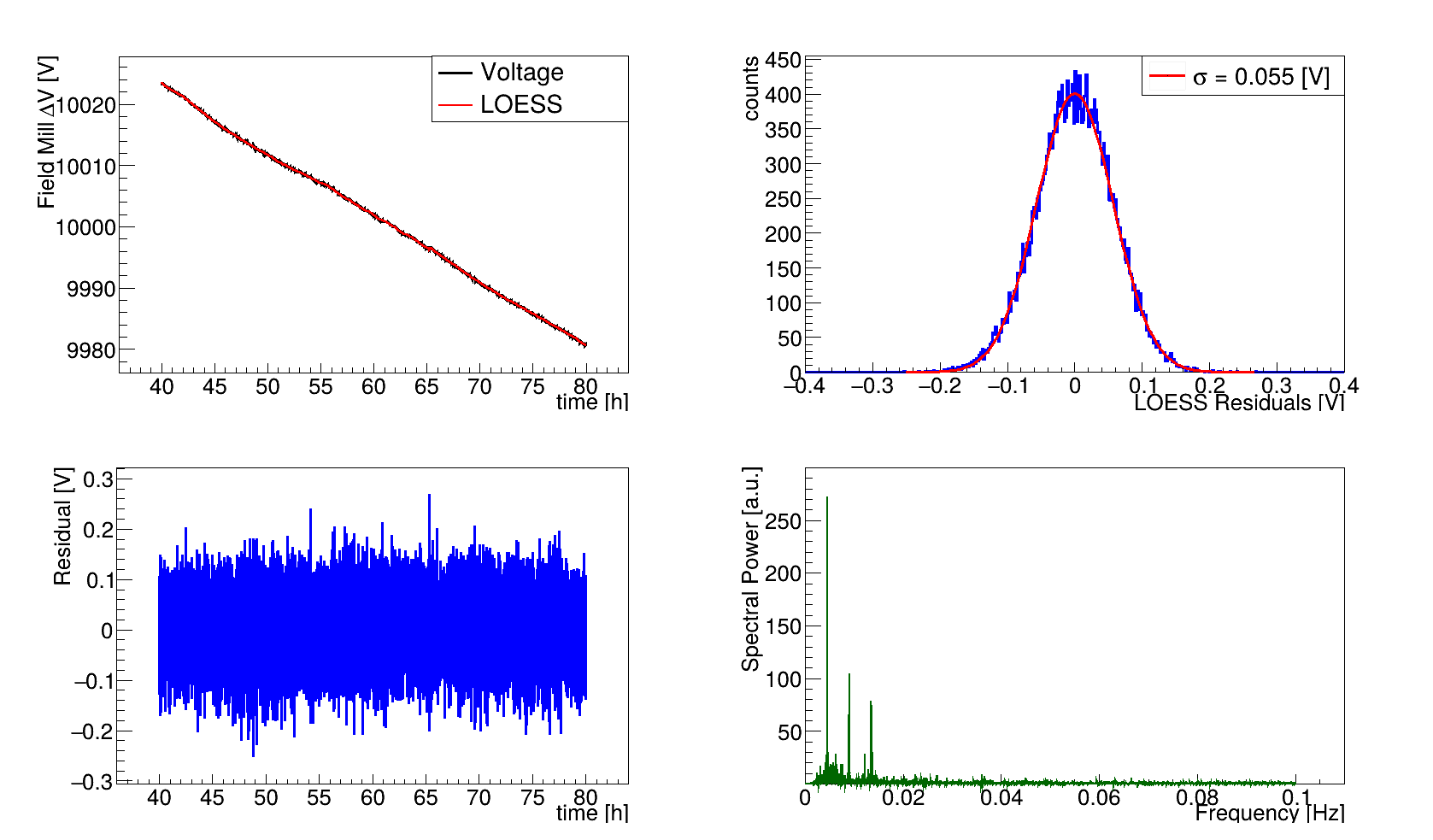} \caption{\emph{Top left:} 10-board chain voltage as a function of time measured by the \emph{field mill} device read out by the $7\sfrac12$-digit precision multimeter. The over-imposed \emph{red} curve represents the local regression. \emph{Top right:} Gaussian distribution of the residuals (in \emph{bottom left}) after LOESS detrend. \emph{Bottom Right:} DFFT of the time series, showing some low frequency periodic component in the time series.} 
\label{fig:V2} \end{figure}
By dividing the time series of the measured voltage residuals $X[i] = V_1(t_i)$ and $Y[i] = V_3(t_i)$, corresponding to the two voltages above, into $N_{\rm ck}$ chunks of 128 sampling points, namely $X_j[i]$ and $Y_j[i]$,  the phase coherence for each frequency $\omega$ is defined as
\begin{equation}
\gamma(\omega) = \frac{1}{N_{\rm ck}} \sum_{j = 1}^{N_{\rm ck}} \frac{S_{xy}^{(j)}(\omega)}{S_{xx}^{(j)}(\omega) S_{yy}^{(j)}(\omega)}\, ,
\end{equation}
where $S_{xy}^{(j)}$ is the cross-spectral density between $X_j[i]$ and $Y_j[i]$, and $S_{xx}^{(j)}$ and $S_{yy}^{(j)}$ are the (Fourier) spectral densities of $X_j[i]$ and $Y_j[i]$ respectively, done through a DFFT algorithm, respectively. By definition, this quantity ranges in $[-1, 1]$. Alternatively, the synchronization, \emph{i.e.} the correlation computed chunk by chunk, which also ranges in the same interval, is defined as
\begin{equation}
\rho_j = \frac{{\rm Cov}(X_j[i], Y_j[i])}{\sigma_{X_j[i]} \sigma_{Y_j[i]}}\, ,
\end{equation}
that is, the covariance between $X_j[i]$ and $Y_j[i]$, normalized to their RMS values.

In Fig.~\ref{fig:corr} we report the phase coherence and synchronization between $V_{\rm 1kV}$ and $V_{\rm 0.5kV}$ over the time interval 3–18~h, corresponding to a highly stable dataset. The plots show a nearly constant value around 0.5 for both $\gamma(\omega)$ and $\rho_i$, exactly as expected from the number of REFs involved in the overlap (50 out of 100), and without any significant correlation in either phase or time. This supports the optimistic approximation in Eq.~(\ref{eq:sim}).  

It is worth pointing out here that precision multimeters cannot measure voltages greater than 1 kV, whereas the 20 kV of the full diode chain can only be monitored by the field mill. In the case of field mill data, however, the situation is a bit worse. In fact, from the measurements conducted over half the chain (10 boards totaling \SI{10}{kV}), the precision, as reported in Fig~\ref{fig:V2}, is $\sigma_{10} \simeq 50$ \unit{\milli V}, corresponding to about 5 ppm, rather than the optimistic $\approx$ \SI{1}{mV} coming from the extrapolation in Eq.~(\ref{eq:sim}). Since in the 10-board setup all additional noise sources are already included, it is not entirely incorrect, in this case, to sum in quadrature the precision of the 10-board setup to estimate the precision of the full REF chain. In this less optimistic case, the total precision will be $\sim 70$ mV, corresponding to $\sim 3.5$ ppm. In this scenario, the expected contribution to the energy resolution will be, according to Eq.~(\ref{eq:propV}), $\sigma_{\rm HV}(E)\simeq$\SI{70}{meV}. This value would be slightly critical for the C$\nu$B detection but fully satisfactory for the neutrino mass measurement.

The origin of the extra noise is still under investigation. According to the tests conducted at LNGS, it is unlikely to be due to correlated noise among the boards, as no such correlation is observed when summing every combination of board sub-parts. Instead, it is more likely due to the \emph{field mill} stability and grounding. This explanation is partially supported by the low noise periodic components in the DFFT visible in Fig.~\ref{fig:V2} \emph{bottom right}. Furthermore, there was substantial evidence of changes in noise conditions when modifying, \emph{e.g.}, the motor type, the motor driver encoding, the ball bearing resistivity, the mill thickness, and so on. 

\section{Fast switch concept}
\label{sec:fast}

\begin{figure} [h!]
    \centering
    \includegraphics[width=\linewidth]{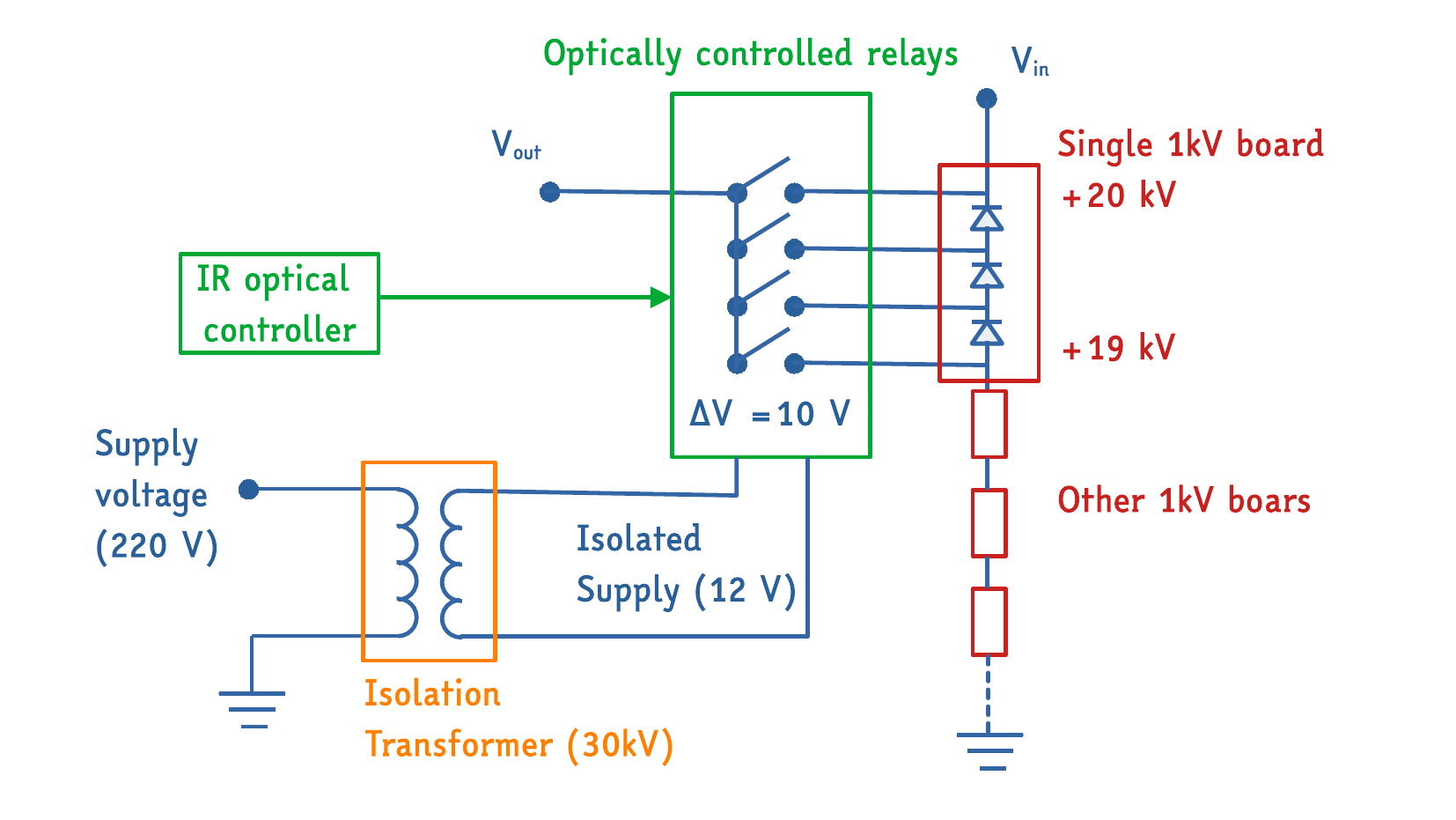}
    \caption{\emph{Fast switch} block diagram. Each 1 kV board can be partitioned with optically controlled high voltage relays. The microcontroller operating voltage is supplied by a isolated transformer.}
    \label{fig:relais}
\end{figure}

As discussed in Sec.~\ref{sec:demo}, to select electrons with the energy on the region of interest close to the $\beta$-spectrum end-point,  the electric fields in the novel EM dynamic filter have to be activated on a short time scale of the order of a few \unit{ms}. This can be realized by a suitable partitioning of the \SI{20}{kV} REF chain (2000 steps of \SI{10}{V} each) activated by high-voltage low-noise relays. 
In order not to disrupt the exceptional stability provided by the diode chain, one requires careful design to maintain precision, minimize leakage currents, and ensure long-term stability, 

A block diagram of the fast switch under design is shown in Fig.~\ref{fig:relais}. Each of the 1~kV boards (highlighted in \emph{red}) can supply the desired voltage in steps of $\Delta V = 10~\mathrm{V}$ by diverting the corresponding switch through optically controlled relays. The microcontroller (in \emph{green}) managing the switch chain is powered by typical low voltages ($12 \div 24$~V) through an isolation transformer rated up to \SI{30}{\kilo\volt} for safety reasons. This module can be replicated across the 20 boards to achieve a full dynamic range from 0 to 20~kV.

Given the high-voltage nature of the system, relays must meet strict requirements. They must be capable of handling at least 1 kV range while ensuring low leakage currents to preserve precision, low contact resistance to avoid voltage drops, and high insulation resistance to minimize unwanted leakage paths. Additionally, if power efficiency is a concern, latching relays may be preferable to reduce energy consumption in steady-state conditions. The most suitable relay types for this application include electromechanical relays (EMRs) for their robustness, solid-state relays (SSRs) for fast and reliable switching, and high-voltage reed relays, which are particularly effective in precision circuits due to their minimal leakage and stable characteristics.  

The relays should be integrated into a circuit topology that allows dynamic segmentation of the voltage reference chain.  
To efficiently manage relay switching, a dedicated control system is required. A microcontroller or FPGA can be used to trigger relays, with isolated gate drivers ensuring safe operation in a high-voltage environment.  
 Using shielded wiring will help reduce electromagnetic interference (EMI), while thermal stabilization of the system will prevent unwanted voltage drift caused by temperature variations. Additionally, selecting relays with low contact resistance and ensuring proper grounding will prevent fluctuations in the measured voltage.  

\begin{figure}
    \centering
    \includegraphics[width=0.8\linewidth]{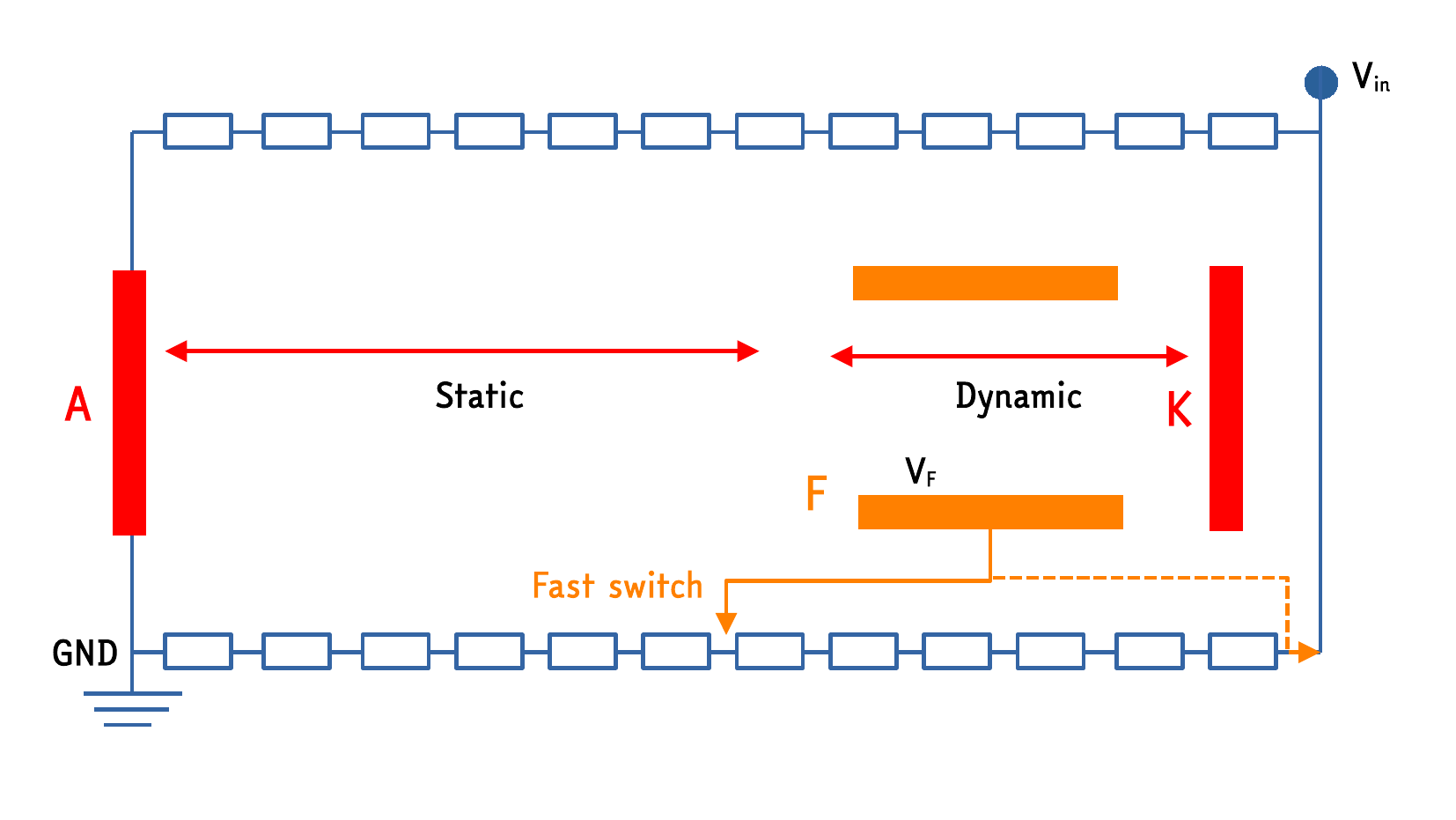}
    \caption{Block diagram of the dynamic filter configuration with the fast switch. The fast switch (F) is inserted in between the anode (A) and the cathode (K) in a dynamic point in the REF chain. Two REF chains are providing a constant maximal voltage from A and K and a dynamic voltage between F and A.}
    \label{fig:fastHV}
\end{figure}

Figure~\ref{fig:fastHV} shows the block diagram of the filter with the fast switch (F) placed in between the anode (A) and the cathode (K) at a dynamic point (\emph{orange arrow}) along the bottom REF chain. The fast switch enables the dynamic filters electrodes that conveys the electrons with energy in the region of interest from to source (A) towards the high-precision calorimeters (K). The top chain represents the maximum voltage on A (\emph{i.e.} $V_{\rm in} = 18.6~\mathrm{kV}$ for the tritium energy endpoint) and corresponds, \emph{e.g.}, to the same voltage applied to K in the first part of the filter (\emph{orange solid dashed}), when the electron is drifting in the constant-$B$ region (CRES). Once the electron enters the dynamic region, the voltage is rapidly switched to the desired value $V_{\rm F}$ after the electron momentum components have been estimated in the previous stage.
It is worth noting that, regardless of the behavior during the transient, the final contribution to the total electron energy due to the electric field is determined solely by $V_{\rm in}$, which is highly accurate. Indeed, the ``$\mathbf{E} \times \mathbf{B}$'' drift in the constant-$B$ field does no work and therefore does not subtract energy from the electrons. Instead, the accuracy of $V_{\rm F}$ only affects the filter efficiency, \emph{i.e.}, the capability of the filter to select a narrow energy region close to the $\beta$-spectrum end-point. In practice, this means that the fast-switch transient and its intrinsic roughness do not significantly affect the total contribution to the energy uncertainty, which is ultimately determined by the accuracy of the reference voltage.

Before implementing the full 2000-unit system, it is essential to build and test a small-scale prototype with 10 to 100 REF modules (\SI{1}{kV}). This allows validation of the relay switching behavior, stability of the output voltage, and measurement of relay leakage currents to ensure they remain within acceptable limits. A prototype is being realized at LNGS in cooperation with the electronics laboratory. The final goal is to prove the preservation of the precision featured in the static measurement,  
the readiness of activation on very short time scales,  
and the absence of possible disturbances in terms of electromagnetic noise interfering with the PTOLEMY dynamic filter performance.

\section*{Conclusions}

We have presented, for the first time, the high-voltage (HV) system developed for the electrodes of the PTOLEMY filter. The system, consisting of a chain of \SI{10}{V} high-precision voltage references arranged in 20 boards of \SI{1}{kV} read out by a \emph{field mill} device interfaced with precision multimeters, has been demonstrated to achieve an impressive precision of $\sim$50 mV over a 10 kV range, which, when extrapolated, translates to an energy resolution contribution of about 70 mV at the tritium end-point (3.5 ppm precision). This level of precision is critical for the detection of C$\nu$B, while the current value meets the necessary requirements for a competitive small-scale experiment aimed at measuring the neutrino mass. The time instability of the HV system consists of slow drifts on a time scale longer than the transit time of the electron in the PTOLEMY filter. Those drifts, as shown in Sec.~\ref{sec:res}, can be easily removed by a \emph{detrend} procedure, and the voltage can be calibrated on the true value via a precision-multimeter in real time.

Considering the intrinsic precision of each \SI{1}{kV} board (in the worst case 0.2 ppm), there is plenty of room for improvement, up to $\lesssim 0.05$ ppm from the uncorrelated noise extrapolation. As indicated by different tests, the extra contribution to the noise reducing the performance potentially originates from the mechanical stability, grounding, and the driver of the \emph{field mill} motor plus blade system, while the low-noise amplification system and readout, according to independent measurements, are unlikely to contribute to the reduction in precision.  

Finally, a \emph{fast switch} implementation for the dynamic constant drift filter is under investigation. The PTOLEMY Collaboration is actively developing prototypes involving low-noise high-voltage relays capable of providing great stability and precision, as demonstrated in the static configuration.  

\section*{Acknowledgments}
A. Colijn is supported by the \emph{Dutch Research Council} (NWA.1292.19.231). C. Tully is supported by the \emph{John Templeton Foundation} (No. 62313). This work is also supported by PRIN grant ``ANDROMeDa'' (PRIN$\_$2020Y2JMP5) of \emph{Ministero dell’Università e della Ricerca}. Finally, the authors acknowledge the support of the INFN CSN-V.

\end{document}